\documentclass[11pt]{article}
\usepackage{cospar}
\usepackage{url}
\usepackage{txfonts}

\usepackage{graphicx}
\usepackage[figuresright]{rotating}

\usepackage{natbib}
\usepackage{amssymb}

\newcommand\psr{PSR J0218$+$4232}
\newcommand\psra{PSR B1821-24}
\newcommand\psrb{PSR B1937+21}

\newcommand\arcdeg{\mbox{$^\circ$}}%
\newcommand\arcsec{\mbox{$^{\prime\prime}$}}%
\newcommand\fdg{\mbox{$.\!\!^\circ$}}%
\newcommand\farcs{\mbox{$.\!\!^{\prime\prime}$}}%
\title{CHANDRA AND RXTE STUDIES OF THE X-RAY/$\gamma$-RAY MILLISECOND PULSAR
\mbox{\psr}}

\author{L. Kuiper\address{SRON National Institute for Space Research, Sorbonnelaan 2, 3584 CA Utrecht, The Netherlands},
	W. Hermsen$^1$, and B. Stappers\address{ASTRON, Postbus 2, 7990 AA Dwingeloo, The Netherlands} }
	
\begin{document}

\maketitle

\begin{abstract}
We report on high-resolution spatial and timing observations of the millisecond pulsar
\psr\ performed with the Chandra X-ray Observatory (CXO) and the Rossi X-ray 
Timing Explorer (RXTE). With these observations we were able to study {\em a\/}) the 
possible spatial extent at X-ray energies of the DC source coincident with \psr\ in 
detail (CXO), {\em b\/}) the relative phasing between the X-ray, radio and $\gamma$-ray 
profiles (CXO and RXTE) and {\em c\/}) the spectral properties at energies beyond 10 keV (RXTE).
We found no indications for extended emission at X-ray energies down to 
$\sim 1^{\prime \prime}$ scales and confirmed the presence of a point-like DC-component. 
The 2 non-thermal pulses in the X-ray profile are found to be aligned with 2 of the 3 
pulses visible at radio-frequencies and more importantly with the two $\gamma$-ray pulses 
seen in the EGRET 100-1000 MeV pulse profile. The latter reduces now the random occurrence probability 
for the detected $\gamma$-ray signal to $\sim 10^{-6}$, which corresponds to a $4.9\sigma$ 
detection significance. This strenghtens the credibility of our earlier claimed detection of 
pulsed high-energy $\gamma$-ray emission from this millisecond pulsar. The pulsed RXTE spectrum 
appears to extend to $\sim 20$ keV and can be described by a power-law with a photon index of 1.14, 
slightly softer than measured by BeppoSAX and XMM for energies below 10 keV. We compare the high-energy
characteristics of \psr\ with those of two other X-ray emitting millisecond pulsars \psra\ and \psrb.
 
\end{abstract}

\section*{INTRODUCTION}

Millisecond pulsar \psr\ is remarkable in many respects. It has a very broad and complex radio pulse
profile with 3 pulses (\citealt{navarro}, \citealt{stairs}), and a $13\pm9\%$ radio DC contribution 
(\citealt{kuiper2002}), suggesting that we continously look into the radio beam. 
At X-ray energies a double peaked profile has been observed (ROSAT HRI: \citealt{kuiper1998}; BeppoSAX 
MECS: \citealt{mineo}; Chandra HRC-S: \citealt{kuiper2002}), and also in this regime a DC-component 
has been detected, significantly only at soft X-rays, and possibly originating from a spatially extended  
region (\citealt{kuiper1998}).
The pulsed 0.1-10 keV X-ray spectrum turns out to be very hard with a photon index in the range 0.5 -- 1.0.
Surprisingly, marginally significant pulsations have also been reported at high-energy ($>100$ MeV)
$\gamma$-rays from a sky region containing an $\sim 11\sigma$ EGRET source, 3EG J0222+4253, which is 
positionally consistent with \psr\ (\citealt{kuiper2000}). 
The 100-1000 MeV $\gamma$-ray profile shows one prominent pulse and a weaker broader one,  
coincident with 2 of the 3 radio-pulses. The high-energy $\gamma$-ray emission is consistent with being 
100\% pulsed and its spectrum can be described by a power-law with a photon index $\sim 2.6$.
Unfortunately, inaccuracies in the absolute timing of the events during the ROSAT and BeppoSAX 
observations made direct comparisons of the soft and medium energy X-ray profiles with the radio and 
high-energy $\gamma$-ray profile impossible.
The aim of the Chandra and RXTE observations was to shed light on the spatial extent of the
X-ray counterpart of \psr\ (CXO), to obtain high-resolution X-ray pulse profiles which could be phase 
related to the radio and $\gamma$-ray profiles (CXO and RXTE), and to extend the spectral coverage 
to energies beyond 10 keV (RXTE).

\section*{CHANDRA AND RXTE RESULTS}

\subsection*{CHANDRA Observations and Results}


\begin{figure}[t]
  \vspace{-1.25cm}
  \begin{minipage}[t]{8.85cm}
    \hspace{0.2cm} \vspace{0.0cm} 
    \includegraphics[width=9cm,height=10cm]{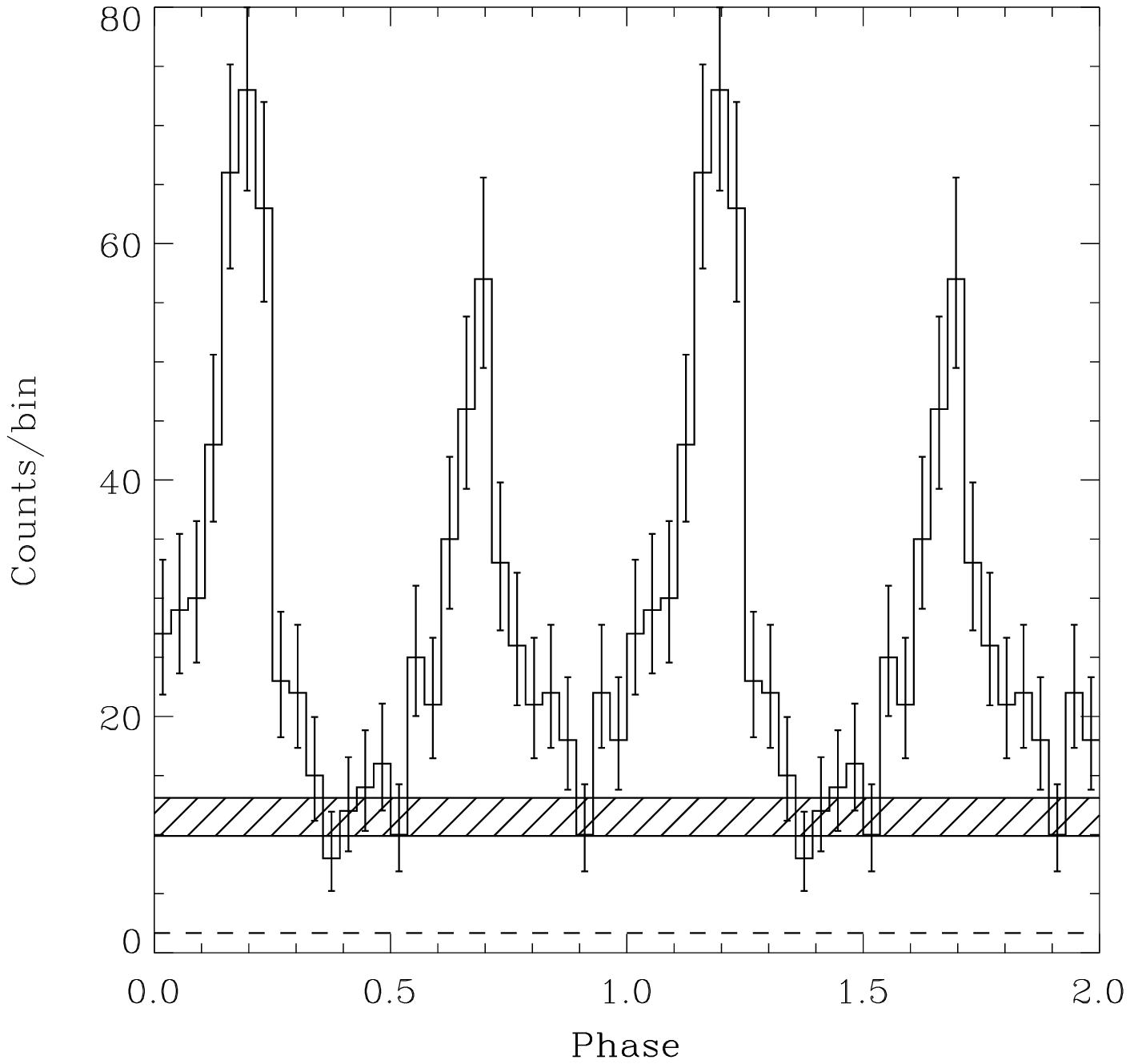}
    \vspace{-0.75cm}
    {\caption{\label{hrcs_profile}
               \ Pulse profile of \psr\ (0.08-10 keV) as observed by Chandra HRC-S in imaging
                mode. Two cycles are shown for clarity. Typical error bars are indicated.
                The background level from the spatial analysis is shown as a dashed line,
                while the hatched area indicates the DC-level $\pm 1\sigma$. The DC-fraction 
                is $0.36\pm0.06$. (see also {Kuiper et al. 2002}) 
                }}
  \end{minipage}
  \hspace{0.1cm}
  \begin{minipage}[t]{8.85cm}
    \hspace{0.5cm} \vspace{0.0cm} 
    \includegraphics[width=9cm,height=10cm]{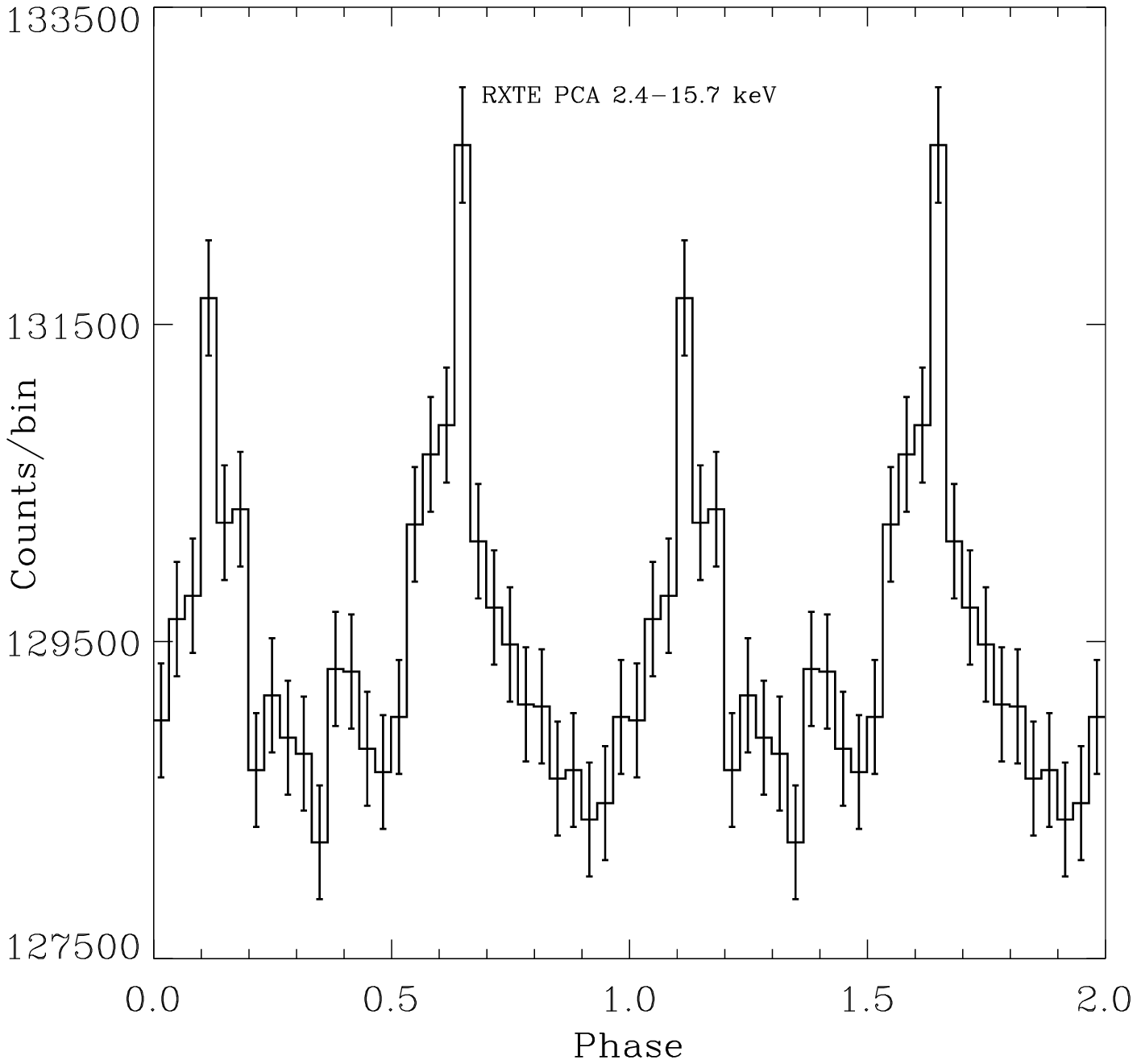}
    \vspace{-0.75cm}
    {\caption{\ Pulse profile of \psr\ as observed by the RXTE PCA in the 2-16 keV energy 
                window (30 bins; only data from the PCA top layer are used). Two cycles are 
                shown for clarity. Error bars indicated are $1\sigma$. 
                The modulation significance is $\sim 12.5\sigma$ applying a $Z_4^2$ test. 
                Notice the morphology change relative to the soft X-ray profile for the Chandra HRC-S 
                shown in Figure \ref{hrcs_profile}. \label{pca_profile}}}
  \end{minipage}
  \vspace{-0.45cm}
\end{figure}


Chandra observed \psr\ twice with the High Resolution Cameras HRC-I and HRC-S. These cameras
are multichannel plate detectors sensitive to X-rays in the 0.08-10 keV range with no spectral resolution.
The first observation on 22-12-1999 with the HRC-I in the focal plane of the telescope mirror and the 
second observation on 5-10-2000 with the HRC-S in the special imaging mode yielded effective exposures 
of 74.11 ks and 73.21 ks, respectively. \psr\ was detected significantly ($\sim 85\sigma$) in the first 
observation. Its spatial extent is consistent with that of a point-source. Thus we have no evidence of 
extended X-ray emission from \psr\ at scales larger than $\sim 1\arcsec$.

The second HRC-S observation was devoted to the timing analysis of \psr. Phase folding the barycentered
times of the X-ray events falling within $1\farcs 5$ from the centroid of the X-ray counterpart of \psr\ 
yielded the well-known double peaked profile with high statistics (see Figure \ref{hrcs_profile}).
The deviation from being flat applying a $Z_6^2$-test is $15.2\sigma$. Indicated in Figure \ref{hrcs_profile}
as a hatched area is the DC or unpulsed level ($\pm 1\sigma$) determined using a bootstrap method 
(\citealt{swanepoel}). Also indicated as a dashed line is the background level derived from an imaging 
study. The pointlike DC component can clearly be discerned in Figure \ref{hrcs_profile}.
The fraction of source counts over the entire 0.08-10 keV energy range from this DC-component is 
$0.36\pm0.06$. Combining the Chandra HRC-S DC count rate with those observed by the ROSAT HRI and 
BeppoSAX MECS and assuming a spectral model for the DC emission we can put constraints on the spectral
parameters. For a power-law model the $1\sigma$-range for the photon index is 1.3--1.85, much softer
than the pulsed emission (\citealt{kuiper2002}). 

Chandra's absolute timing accuracy is within $200\mu$s (\citealt{tennant}) allowing detailed phase
comparisons with the radio and $\gamma$-ray profiles. It turns out that the two non-thermal X-ray pulses 
are aligned with the two $\gamma$-ray pulses and two of the three radio pulses (see also the section about the {\it Rossi\/} 
X-ray Timing Explorer (RXTE) PCA/HEXTE timing analyses and Figure \ref{multiprof}) given the timing inaccuracies of the instruments. 
The random occurrence probability that we 
are dealing with a 100-1000 MeV $\gamma$-ray profile with a $3.5\sigma$ modulation significance {\em and}
alignment between the X-ray and $\gamma$-ray pulses decreases to about $10^{-6}$ i.e. the significance
of the detection of the pulsed $\gamma$-ray signal from \psr\ improves to $\sim 4.9\sigma$. For more details on the Chandra results, see \cite{kuiper2002}. 

\subsection*{RXTE Observation and Results}

RXTE observed \psr\ for about 200 ks between 26-12-2001 and 7-01-2002 (MJD time span 
52269 -- 52281). The Proportional Counter Array PCA (2-60 keV) data were obtained in the Good Xenon mode, time tagging 
each event with a $0.9\mu$s time resolution. The two detector clusters of the High Energy X-ray Timing Experiment HEXTE (10-250 keV) 
operated in staring mode, and the science mode was {\small \sf  E\_8us\_256\_DX1F} allowing 
spectroscopic studies with 256 channels and a time tag resolution of $7.6 \mu$s. 
During the observation period \psr\ was regularly monitored by the Dutch pulsar machine PuMa 
(\citealt{voute}) installed at the Westerbork Synthesis Radio Telescope (WSRT) allowing 
high precision timing studies.

\subsubsection*{RXTE PCA/HEXTE Timing Analyses}
\label{rxte_timing}
In the timing analysis of the PCA data the barycentered event times were folded with the
newly obtained WSRT PuMa ephemeris (MJD range: 51386 -- 52278) as well as with an earlier 
ephemeris based on timing measurements at Jodrell Bank (MJD range: 49092 -- 51462). No 
significant differences in the X-ray pulse morphologies were seen folding the events with the
different ephemerides.
 
The modulation significance in the $2.4$-$15.7$ keV window is $12.5\sigma$ applying a $Z_4^2$-test:
a 30 bin pulse profile is shown in Figure \ref{pca_profile}. The two X-ray pulses are located at approximately the same phases as in the CXO HRC-S lightcurve, however, the relative strengths of 
the pulses have changed.
The pulsed signal can be detected up to $\sim 20$ keV. The significances in the $2.4$-$7.8$
keV and $7.8$-$15.7$ keV windows are $10.6$ and $6.0\sigma$, respectively, applying again
a $Z_4^2$-test. Therefore, for the first time significant pulsed emission has been detected 
at energies beyond $\sim 8$ keV.

\begin{figure}[t]
  \vspace{-1.25cm}
  \begin{minipage}[t]{8.85cm}
    \hspace{-0.5cm} \vspace{0.0cm} 
    \includegraphics[width=10cm,height=15cm]{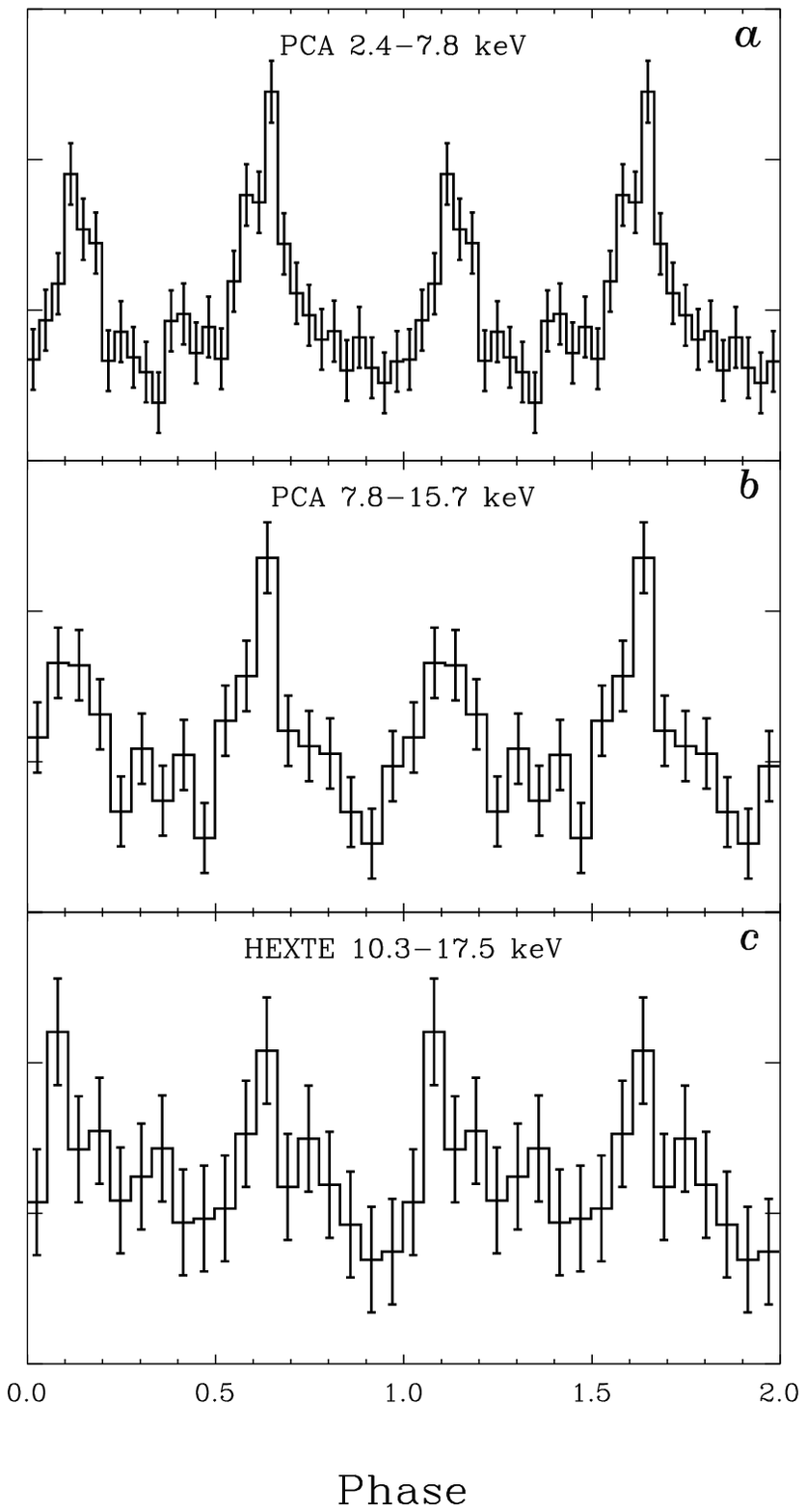}
    \vspace{-0.95cm}
    {\caption{ Pulse profiles of \psr\ as measured by the RXTE PCA for the 2.4-7.8 and
    7.8-15.7 keV energy windows and RXTE HEXTE for the 10.3-17.5 keV energy window. 
    \label{pca_hexte_profiles}}}
  \end{minipage}
  \hspace{0.1cm}
  \begin{minipage}[t]{8.85cm}
    \hspace{0.5cm} \vspace{0.0cm} 
    \includegraphics[width=8.95cm,height=15cm]{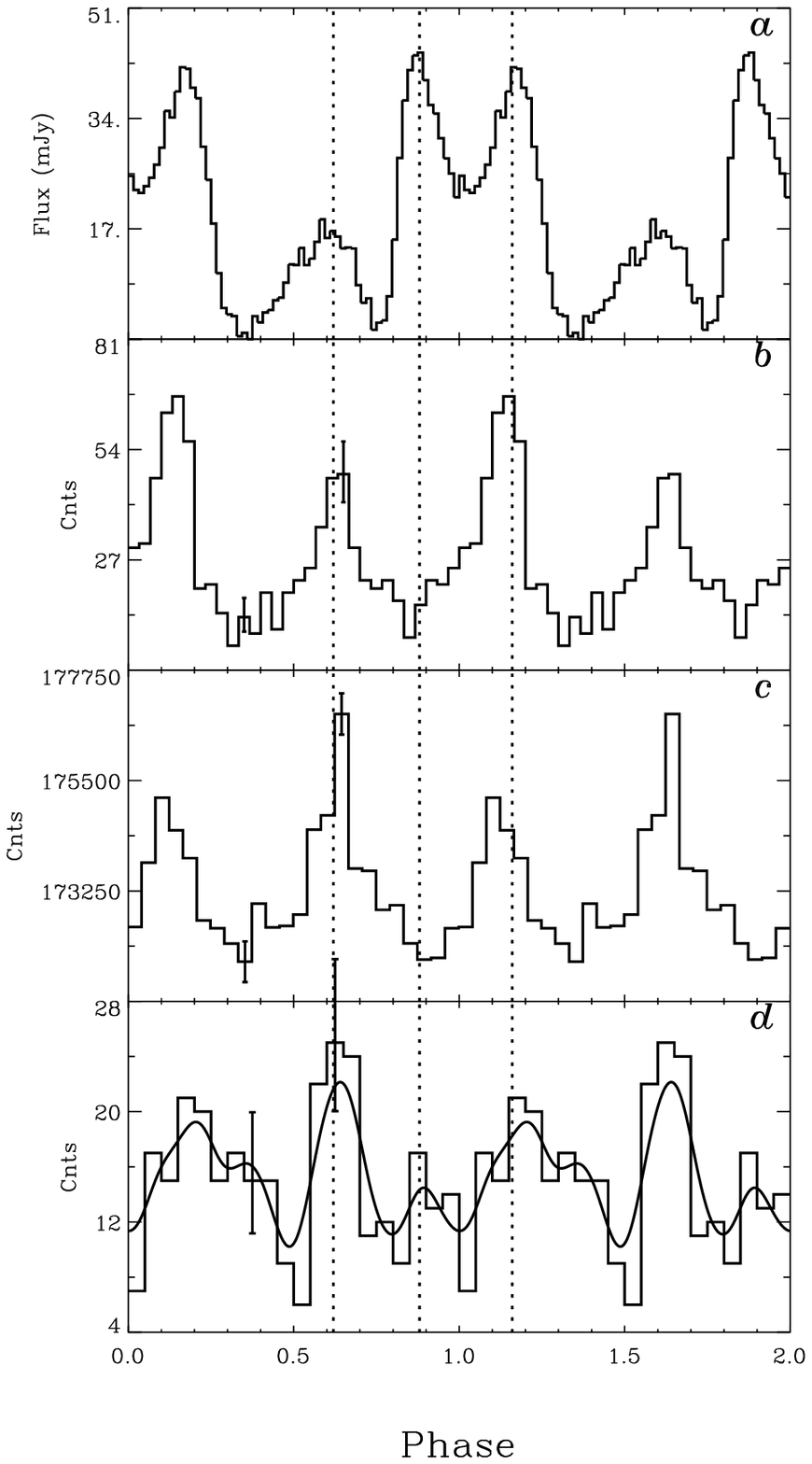}
    \vspace{-0.95cm}
    {\caption{ Multi-wavelengths pulse profiles of \psr\ in absolute phase.
    From top to bottom: Radio: 610 MHz, X-rays: Chandra HRC-S 0.08-10 keV
    and RXTE PCA 2-16 keV, and $\gamma$-rays: CGRO EGRET 0.1-1 GeV.
    \label{multiprof}}}
  \end{minipage}
  \vspace{-0.45cm}
\end{figure}

The HEXTE data were also screened using timeline filters. Pulse-phase folding the event times 
from the combined Cluster A + B screened datasets yielded pulse-phase distributions in 256 PHA channels.
The pulsed signal has been detected in the lower PHA channels 9 -- 16, corresponding to 10.3 -- 17.5 keV, 
reaching a detection significance of $\sim 3.7\sigma$. A compilation of PCA/HEXTE pulse profiles in 
three different energy windows is shown in Figure \ref{pca_hexte_profiles}.

It is now also possible to cross-correlate the Chandra and RXTE profiles in the overlapping 
energy range. Taking into account the internal delays (PCA 16$\mu$s; Chandra HRC-S 19.5$\mu$s) 
we found that the Chandra clock runs $105\pm 17\mu$s ahead the RXTE clock. This is a significant 
improvement upon a previous estimate based on a Chandra-RXTE correlation study of the much slower 
rotating Crab pulsar, for which absolute timing analysis is further complicated because of 
irregular Dispersion Measure variations.
The Chandra phases have therefore been adjusted to the RXTE profile by applying a back-shift
of 0.045.  
Figure \ref{multiprof}, finally, shows the comparison in absolute phase of the \psr\ Chandra HRC-S 
and RXTE PCA X-ray profiles with the radio and $\gamma$-ray profiles. 
The non-thermal X-ray pulses appear to be aligned within the timing uncertainties with two of the 
three radio pulses and with the two $\gamma$-ray pulses.


\subsubsection*{RXTE PCA/HEXTE Spectral Analyses}

In the spectral analysis of the PCA data we treated the 5 Proportional Counter Units (PCU's) constituting the PCA separately.
Effective exposure times for the individual PCU's are determined excluding time periods during which
an increased background is encountered e.g. during a South Atlantic Anomaly passage. Moreover, time intervals are ignored
when the angle between target and Earth Horizon is less than $5\arcdeg$ or the target aspect angle is larger 
than $0\fdg 05$. We further considered only event triggers from the top Xenon layer. 
Response matrices and sensitive areas were derived for each PCU separately using the {\sf ftools} (v5.2, 
release 18-6-2002) program {\sf pcarsp} (v8.0) and {\sf xtefilt} (v1.7).

Next, for each individual PCU pulse-phase histograms have been generated from the screened PCU event 
datasets in 256 Pulse Height Analyzer (PHA) channels. In order to determine the pulsed counts from \psr\ in different PHA
slices we first made a profile template based on a fit of the total 2-16 keV pulse profile in terms of two
asymmetric Lorentzians atop a flat background. The two asymmetric Lorentzians (fixed shape and position)
were fitted, each with free normalization, atop a flat background with a free scale to the measured 
pulse-phase distribution for each selected PHA interval. The sum of the two normalizations constitutes then the total number of pulsed counts per PHA interval. 

\begin{figure}[t]
\vspace{-0.45cm}
\hspace{4cm} \vspace{0.0cm}
 \includegraphics[width=10cm,height=10cm]{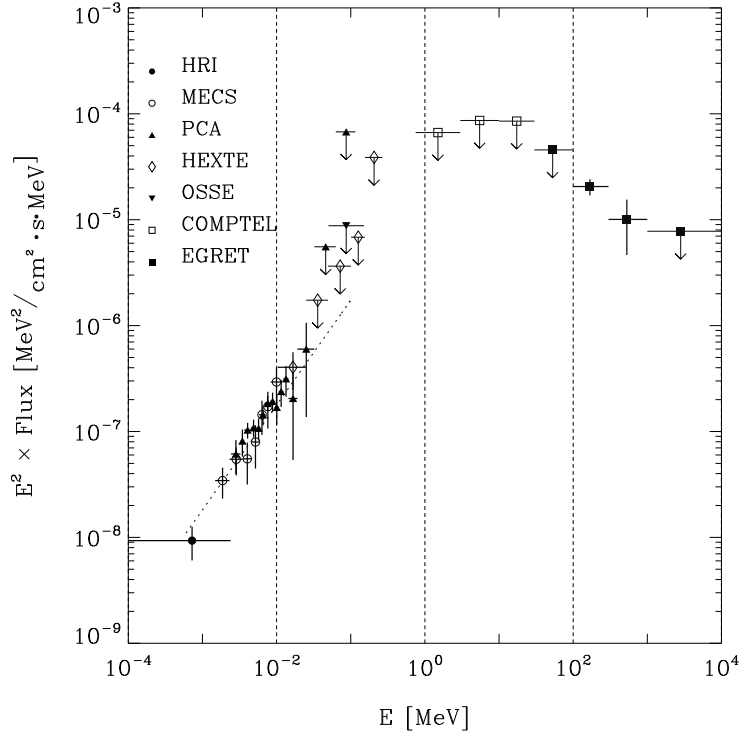}
\vspace{-0.5cm}
\caption[h]{\label{he_spectrum}\ \ Pulsed high-energy spectrum of \psr\ in a $\nu\times F_{\nu}$ 
     representation from soft X-rays up to high-energy $\gamma$-rays including the newly derived
     RXTE PCA/HEXTE and revisited BeppoSAX flux measurements and a spectral fit (dotted line; 
     extrapolated up to 100 keV) based on XMM EPIC-PN total pulsed data (fit range: 0.6-10 keV; 
     Webb, 2003).
}
\end{figure}

A simple absorbed power-law was fitted to the data using a fixed N$_{\hbox{\tiny \rm H}}$ of $5\cdot 10^{20}$ cm$^{-2}$ 
to the combined PCU pulsed counts, taking into account the individual PCU responses and screened exposures.
The best fit total pulsed spectrum has a photon index of $1.14\pm _{0.04}^{0.03}$ over the full PCA energy
range. Using the best fit PCA based model templates for the two peaks we also revisited the BeppoSAX 
total pulsed emission by applying similar fit procedures in different BeppoSAX MECS energy intervals.
We obtained for the MECS energy range a slightly harder photon index of $0.78\pm _{0.10}^{0.08}$, which 
is a bit different, but consistent with the earlier reported value of 0.61 (\citealt{mineo}). A joint fit of
the revisited BeppoSAX MECS and RXTE PCA spectral measurements yielded a photon-index of 
$0.99\pm 0.03$ over the entire MECS/PCA energy range.
Similar fit techniques as used for the PCA have been employed to determine the number of pulsed counts
for selected HEXTE PHA intervals.
The effective HEXTE Cluster A and B exposures have been determined using {\sf ftools} program 
{\sf hxtdead} (v2.0.0). The actual exposures are typically reduced by 30\% due to deadtime 
effects. Response matrices and sensitive areas have been derived using {\sf ftools} program 
{\sf hxtarf} (v1.9) taking into account aspect offsets between target and pointing.
The HEXTE total pulsed counts for given PHA slices are not fitted through forward folding 
assuming a spectral model in view of the limited source count statistics, but instead 
the count rates are converted to flux values by estimating the effective sensitive area 
in a certain PHA slice assuming a power-law input photon spectrum with index 1. This 
approach yielded one significant flux point for the 10.3-17.5 keV energy window and upper-limits
beyond this range. The newly derived RXTE PCA \& HEXTE and revisited BeppoSAX MECS 
flux measurements are shown in Figure \ref{he_spectrum} along with the ROSAT HRI 
(0.1-2.4 keV) flux point and CGRO OSSE, COMPTEL and EGRET flux values and upper limits (see  
\citealt{kuiper2000}). The PCA flux points are multiplied by 0.81, because the normalization 
of the PCA spectra turns out to be too high by $\sim 20\%$ based on BeppoSAX and RXTE PCA 
cross-calibrations (this work using Crab RXTE PCA data; see also \citealt{kuulkers2002}). 
Superposed as dotted line is also a recent estimate of the pulsed spectrum of \psr\ using 
XMM EPIC-PN data (Webb, 2003), which nicely overlaps with current flux measurements. 

\section{CONCLUSION AND DISCUSSION}

\begin{figure}[t]
\vspace{-0.35cm}
\hspace{-1.35cm} \vspace{0.0cm}
 \includegraphics[width=20.5cm,height=13cm]{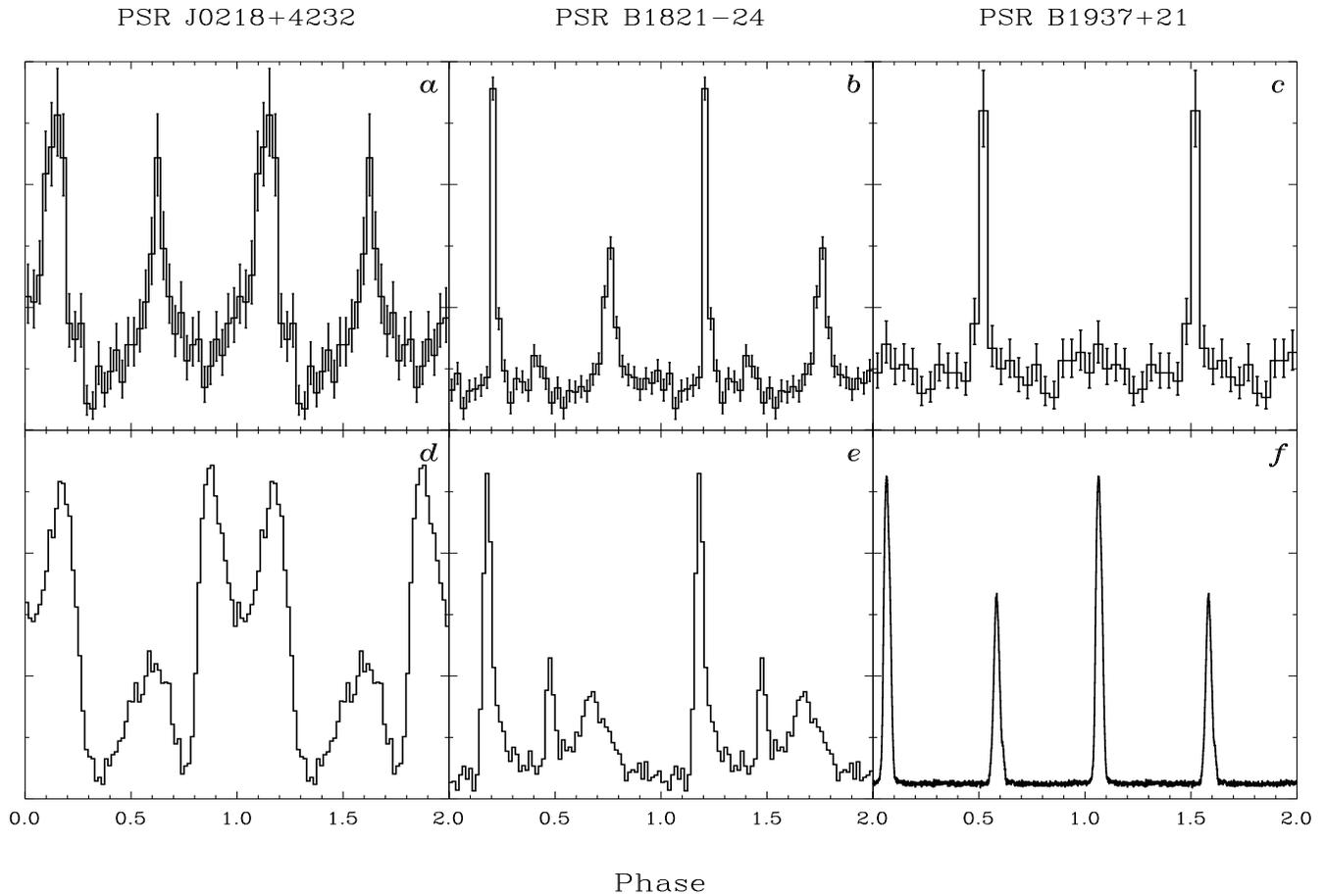}
\vspace{-1.5cm}
\caption[h]{\label{prof_comparison}\ Comparison in absolute phase of the three high-luminosity and hard 
            X-ray millisecond pulsars \psr\ (left; this work), \psra\ (middle; reanalysis of RXTE data) 
            and \psrb\ (right; alignment according to Takahashi et al. 2001) at X-rays (top panels) and 
            radio-frequencies (bottom panels).
}
\end{figure}

From recently performed Chandra and RXTE observations of \psr\ we learned that: {\sf a)} the two 
non-thermal X-ray pulses are aligned with the $\gamma$-ray pulses and with two of the three radio pulses, 
{\sf b)} pulsed emission has been detected significantly up to $\sim 20$ keV, and {\sf c)} the X-ray DC 
component is point-like down to $\sim 1\arcsec$ scales with a (much) softer spectrum than the pulsed 
emission.

How unique is \psr? So far, in X-rays three radio ms-pulsars have been detected showing hard X-ray 
spectra with narrow pulses and having high X-ray luminosities. A comparison in absolute phase of the X-ray 
and radio pulse profiles of this small sample, comprised of \psr, \psra\ and \psrb, is shown in Figure 
\ref{prof_comparison}. Some of the X-ray pulses are clearly aligned with radio pulses, while others are 
only approximately. The geometries i.e. the magnetic inclinations and observer/spin-axis angles are very 
different for these pulsars, implying that we can not draw simple conclusions from the phase comparisons 
and that 3D modeling of the magnetospheric electro-magnetic output is required using the exact geometry to 
explain the high-energy and radio profiles. It can be noted that these three millisecond pulsars are among
the top four ranking millisecond pulsars according to their total spin-down luminosity.

\begin{figure}[t]
\vspace{-0.45cm}
\hspace{4cm} \vspace{0.0cm}
 \includegraphics[width=10cm,height=10cm]{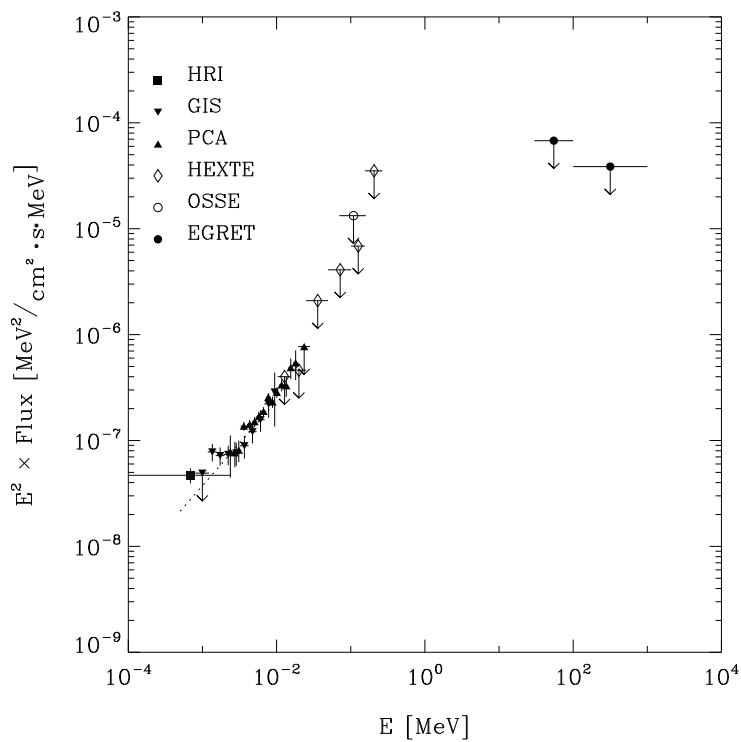}
\vspace{-0.5cm}
\caption[h]{\label{he_spectrum_1821}\ \ Total pulsed high-energy spectrum of \psra\ in a 
   $\nu\times F_{\nu}$ representation from soft X-rays up to high-energy $\gamma$-rays. 
   The ROSAT HRI, ASCA GIS, RXTE PCA and RXTE HEXTE flux values/upper limits are determined 
   in this work. The dotted line shows the total spectrum (0.5 - 8 keV) of \psra\ as 
   derived from CXO ACIS data (Becker et al. 2003). EGRET upper limits are from Fierro et al. (1995).
}
\end{figure}

In the spectral domain the high-energy pulsed spectra of \psra\ and \psrb\ could be very similar to that 
of \psr. For \psra\ we found using archival RXTE PCA/HEXTE and ASCA GIS data that the total pulsed spectrum  
can be described by a power-law with photon index of $1.15\pm 0.02$ for the PCA ($\sim 2$ -- 20 keV) and 
$1.30\pm 0.07$ for the GIS ($\sim 0.8$ -- 10 keV). A joint fit of the ASCA GIS and RXTE PCA data (normalization 
correction of 0.81 applied) yields a photon-index of $1.13\pm 0.02$ assuming an absorbing Hydrogen column of $1.6\cdot 10^{21}$ cm$^{-2}$. The total (i.e. pulsed + DC) spectrum of \psra\ derived recently using CXO 
ACIS data can be described by a power-law with photon index $1.20\pm_{0.13}^{0.15}$ (\citealt{becker03}). 
Both the photon index and normalization of this total spectrum are consistent with the numbers we determined 
for the pulsed component alone, leaving little room for a DC-component. At high-energy $\gamma$-rays
upper-limits are reported (\citealt{fierro}). 
The pulsed high-energy spectrum of \psra\ from soft X-rays up to high-energy $\gamma$-rays is shown in Figure \ref{he_spectrum_1821}. The genuine spectrum could be very similar in shape as shown for \psr\ in Figure 
\ref{he_spectrum}.

For \psrb\ pulsed X-ray emission was detected by \cite{takahashi}. \cite{nicastro} measured a photon index 
of $1.71\pm 0.07$ for the total spectrum for energies up to 10 keV, but, sofar, at higher X-ray and gamma-ray energies no detections have been reported. The fact that 
\psr\ is at present the only millisecond pulsar detected at gamma-ray energies can be merely due to the "modest"
sensitivity of previous high-energy instruments for energies above e.g. 30 keV. Another reason might be that the 
present catalog of millisecond radio pulsars is still very incomplete. Namely, part of the unidentified high-energy gamma-ray sources in the COS-B (\citealt{swanenburg}) and EGRET (\citealt{hartman}) catalogues might have as counterparts millisecond pulsars which have been missed sofar in radio searches. The high-energy spectrum measured
for \psr\ (hard X-ray  and soft high-energy gamma-ray spectrum) is indeed reminiscent of that inferred for part of 
these unidentified sources and there are not many other candidate counterparts for the stable unidentified gamma-ray
sources, excluding millisecond pulsars and normal radio pulsars.

Therefore: {\sf 1)} Millisecond pulsars with hard X-ray spectra are good candidates for detection with future 
gamma-ray telescopes (e.g. with the recently launched ESA satellite INTEGRAL, sensitive between 15 keV and 5 MeV, 
and the future Italian mission AGILE and NASA's GLAST, both sensitive to gamma rays with energies above about 30 
MeV); {\sf 2)} Accurate source localization of unidentified gamma-ray sources with future gamma-ray missions might
lead to detections of new millisecond pulsars at radio wavelengths, followed by the identification of the pulsed 
signals in the X-ray and gamma-ray data.

We expect that \psr\ will not remain for long the only millisecond pulsar detected at gamma-ray energies.

\bibliographystyle{natbib}

\end{document}